# Evaluation of Radiation Dose Reduction during CT Scans Using Oxide Bismuth and Nano-Barium Sulfate Shields

**Youl-Hun Seoung, Ph.D**

Department of Radiological Science, Cheongju University Korea, Chungbuk, 390-702

The purpose of the present study was to evaluate radiation dose reduction and image quality during CT scanning by using a new dose reduction fiber sheet (DRFS) with commercially available bismuth shields. These DRFS were composed of nano-barium sulfate ($BaSO_4$), filling the gaps left by the large oxide bismuth ($Bi_2O_3$) particle sizes. The radiation dose was measured five times at directions of 12 o'clock from the center of the polymethyl methacrylate (PMMA) head phantom to calculate an average value using a CT ionization chamber. The image quality measured CT transverse images of the PMMA head phantom depending on X-ray tube voltages and the type of shielding. Two regions of interest in CT transverse images were chosen from the right and left areas under the surface of the PMMA head phantom and from ion chamber holes located at directions of 12 o'clock from the center of the PMMA head phantom. The results of this study showed that the new DRFS shields could reduce dosages to 15.61%, 23.05%, and 22.71% more in 90 kVp, 120 kVp, and 140 kVp, respectively, than with a conventional bismuth shield of the same thickness, while maintaining image quality. In addition, the DRFS were produced to about 25% more thinness than conventional bismuth. We concluded, therefore, that DRFS can replace the conventional bismuth and may be utilized as a new shield.

Email: radimage@naver.com

Fax: +82-43-229-8969

**INTRODUCTION**

The invention of computed tomography (CT) by Godfred Newbold Hounsfield in 1967 revolutionized radiological diagnosis, and the use of routine CT examinations increased rapidly. An estimated 68.7 million CT examinations were performed in 2007 in the United States, with an observed annual growth rate of about 8% since 2003 [1]. Radiation doses for patients have also increased as the number of CT examinations performed continue to increase [2,3]. Recently, patient doses have increased in clinics because of the multi-detector CT (MDCT), which provides quicker images in larger volumes, thus giving larger doses to patients than those from a single-detector CT [4,5]. The International Committee of Radiological Protection (ICRP), in order to achieve its goals, has recommended compliance with radiation protection systems, which require justification for scans, the optimization of protection, and dose limits for radiation protection [6]. In particular, during CT examinations, radiation exposure should be minimized for high radiation sensitive organs such as the lenses of the eyes, the thyroid, the breasts, and the testicles [7]. Therefore radiologic technologists and radiologists must recognize the risks of patient doses during CT examinations and suggest appropriate protocols in order to reduce the doses to which radiosensitive organs are exposed.

Several advanced studies have researched techniques for reducing patient doses during CT examinations. For example, automatic exposure control techniques adjust the tube current along the angular or longitudinal directions, or both, to optimize the dose based on patient size and the scanning parameters [8-11]. Another method uses radio-protective shields, which decrease radiation dose to patients by attenuating photons passing through the shields. General shields of lead can cause streak and beam hardening artifacts below the shield. Bismuth shields are easy to use and have been shown to reduce doses in CT scanning, without artifacts, of anterior organs such as the breast, the lens of the eye, and the thyroid [12-15]. Many research results have been reported in relation to the evaluation of bismuth shields of radiation dose reduction and image quality [16-20]. It has been indicated that bismuth shields make mixing with latex. However, we looked for a new dose shielding, because conventional bismuth has a disadvantage of beam hardening artifact by non-uniform dose distribution in that it adheres poorly to a curved surface such as eyeball, the thyroid, and the breast.

Recently, Kim *et al.* reported a new dose reduction fiber sheet (DRFS) for the scatter X-ray [21]. These DRFS were composed of nano-barium sulfate ($BaSO_4$), filling the gaps left by the large oxide bismuth ($Bi_2O_3$) particle sizes. However, the usefulness of DRFS with respect to image quality and reduction of patient doses during CT examinations has not yet been fully determined when they are placed over the organs of interest. The aim in the present study was to evaluate radiation dose reduction and image quality when using new DRFS with commercially available bismuth shields during CT scanning.

## MATERIALS AND METHODS

### Materials

We performed CT scanning by using a 4-channel MDCT (MX8000-IDT, Philips, Cleveland, OH, USA) scanner installed in Cheongju University. The CT dose was measured by a CT ionization chamber (Model 500-200, Fluke Biomedical, Cleveland, OH, USA) in a polymethyl methacrylate (PMMA) head phantom (Model 76-414, Fluke Biomedical, Cleveland, OH, USA) that had five measurement points at the center and periphery. The CT ionization chamber was a pencil type 3.2 $cm^2$ with sensitive length of 10 cm. The PMMA head phantom was made of solid acrylic, with five cylindrical holes located at the center and at directions of 12, 3, 6 and 9 o'clock from the center. The diameters of the head phantoms were 16 cm. The diameter of the holes was 1.31 cm, and each hole was 1 cm from the edge of the phantom. These locations were close to the eyes and the thyroid [Fig. 1]. We used commercially available bismuth shielding for thyroid by F & L Medical Products Co. (Vandergrift, PA, USA). The density of the bismuth shielding was approximately 0.7 $g/cm^{-3}$. In order to produce sheet forms, the DFRS consisted of the radiation shielding coated with layers and fabrics. The composition of the radiation shielding coating layer consisted of an average particle size of 1–500 ㎛ of bismuth oxide power in the liquid silicone rubber base and an average particle size of 5–50 nm nano-particles of barium sulfate [22]. We coated and hardened the radiation shielding coating layer on one side of fabric and manufactured the DRFS to thicknesses of 0.15 mm, 0.25 mm, 1.00 mm, and 2.00 mm [Fig 2]. Figure 3 is a cross-sectional schematic diagram of the DRFS. Although the shapes of these shields were different in the study, it didn't affect to radiation dose and image quality because we performed scans a single of cross-sectional image of shields on the same phantom.

**Scanning conditions**

Scanning conditions were defined by various parameters such as the number of channels, slice thickness (mm), tube current (mAs), scanning time (sec/rotation), X-ray tube voltage (kVp), and scan method. In this study, the parametric values of all the factors except the X-ray tube voltage were fixed. The fixed X-ray exposure factors were 250 mAs, 2 channels, 10 mm slice, 0.75 sec/rotation, and the axial scan method. The X-ray tube voltages were 90 kVp, 120 kVp, and 140 kVp. The radiation dose and the image quality were measured for every combination of the X-ray tube voltages with the other fixed X-ray exposure factors at the PMMA head phantom. As shown in figure 4, the PMMA head phantom was aligned with the center of the CT gantry.

**Measurement of radiation dose**

Each shield was attached around the phantoms as shown in figure 4. The dose was measured five times at directions of 12 o'clock from the center of the PMMA head phantom to calculate an average value. The doses with bismuth shielding and DRFS and without shielding around the PMMA head phantom were compared in order to calculate the dose reduction. The computed tomography dose index (CTDI) is a commonly used radiation exposure index in X-ray computed tomography and is supplied to scan personnel by CT manufacturers for each examination. The CTDI represents the average absorbed dose, along the z-axis, from a series of contiguous irradiations. It is measured from one axial CT scan (one rotation of the X-ray tube) and is calculated by dividing the integrated absorbed dose by the nominal total beam collimation [23-26]. Exposure (R) values measured by the ion chamber and electrometer were converted to absorb in acryl (mGy) as follows (1):

$$CTDI\,(mGy) = \frac{f(mGy/R) \times C \times M(R \cdot cm)}{N \times T(cm)} \quad (1)$$

Where f is the exposure to dose conversion factor (=0.78) in acryl, C is the chamber calibration factor (=0.98), M is the exposure value in monitor (R) × the chamber length factor (10 cm), N is the number of channels, and T is the slice thickness (cm).

**Image quality evaluation**

The measurement of CT number and noise used the CT image analysis program Image J 1.37 ver. (Wayne Rasband, National Institutes of Health, USA). As shown in figure 5, the regions of interest (ROI) of the CT image quality evaluation were 3.0 ± 0.03 mm distance areas under the surface of the PMMA head phantom and right and left areas at a 3.0 ± 0.03 mm distance from the ion chamber holes located at directions of 12 o'clock from the center of the PMMA head phantom. We manually drew the size of the ROI, circles of about 300.0 ± 1.5 mm², in order to keep homogeneity. The mean intensity of the pixels in the ROI were defined CT numbers. To perform noise measurement, the standard deviation, $\delta$, within an ROI of the reconstructed image, $f(i, j)$, is calculated as follows (2):

$$\delta = \sqrt{\frac{\sum_{i,j \in ROI}[f(i,j) - \overline{f}]^2}{N-1}} \quad (2)$$

Where *i* and *j* are indices of the two-dimensional image, N is the total number of pixels inside the ROI, and *f* is the average pixel intensity.

The average CT numbers and noise in each ROI were compared among the shielding methods.

## RESULTS

**Reduction of radiation dose**

Table 1 shows the resulting radiation doses, measured with no shield, the 1 mm bismuth shield, the 0.15 mm DRFS shield, the 0.25 mm DRFS shield, the 1.00 mm DRFS shield, and the 2.00 mm DRFS shield at directions of 12 o'clock from the center of the PMMA head phantom, at 90 kVp, 120 kVp, and 140 kVp. In this study, at higher X-ray tube voltages, the mean radiation dose increased, and the dose reductions by shields decreased. The dose reduction rates of shielding at 90 kVp were 32.20% at 1.00 mm bismuth, 25.63% at 0.15 mm DRFS, 38.02% at 0.25 mm DRFS, 62.81% at 1.00 mm DRFS, and 66.94% at 2.00 mm DRFS. The dose reduction rates of shielding at 120 kVp were 30.02% at 1.00 mm bismuth, 13.56% at 0.15 mm DRFS, 30.85% at 0.25 mm DRFS, 53.07% at 1.00 mm DRFS, and 62.95% at 2.00 mm DRFS. The dose reduction rates of shielding at 140 kVp were 28.10% at 1.00 mm bismuth, 20.26% at 0.15 mm DRFS, 29.72% at 0.25 mm DRFS, 50.81% at 1.00 mm DRFS, and 62.91% at 2.00 mm DRFS. The results of this study show that the reduction of the radiation dose rate was similar between 1.00 mm bismuth and 0.25 mm DRFS in spite of the altered X-ray tube voltages. In addition, the reduction of radiation dose rate was 28% more using the 1.00 mm DRFS and 60% more with the 2.00 mm DRFS.

**Image quality**

Figure 6 shows CT transverse images of the PMMA head phantom using the different shield methods. Two ROIs were chosen for the right and left areas under the surface of the PMMA head phantom and from ion chamber holes located at directions of 12 o'clock from the center of the PMMA head phantom.

Figure 7 (a) shows the results by CT number, where the bars displayed among by means of shield methods in altered kVp. In this study, the mean CT number of every kVp was measured at 127.87 HU with no shield, 235.18 HU with 1.00 mm bismuth, 205.07 HU with 0.15 mm DRFS, 250.56 HU with 0.25 mm DRFS, 642.90 HU with 1.00 mm DRFS, and 1630.38 HU with at 2.00 mm DRFS. The results show that the measured CT number values with no shield were similar to CT number values with the solid acrylic of the AAPM CT phantom. However, the results of this study showed that the CT numbers were similar between 1.00 mm bismuth and 0.25 mm DRFS in spite of altered X-ray tube voltages.

Measured noise results are presented in figure 7 (b). The mean image noise, similar to the evaluation method of the CT numbers, was measured at 7.85 HU with no shield, 74.31 HU with 1.00 mm bismuth, 59.05 HU with 0.15 mm DRFS, 84.12 HU with 0.25 mm DRFS, 268.86 HU with 1.00 mm DRFS, and 791.82 HU with 2.00 mm DRFS. The results particularly show that the CT numbers and noise values rapidly increase because streak artifacts occur in the X-ray images due to a hardening effect at 1.00 mm and more DRFS. However, as shown in figure 7, differences in visual assessment were few in the 1.00 mm bismuth, 0.15 mm DRFS, and 0.25 mm DRFS.

**DISCUSSION**

The purpose of the present study was to place radiation shielding about radiosensitive surface organs to lower the X-ray energy in CT examinations while maintaining the CT image quality. With the development of MDCT, concerns about radiation protection and patient dose in clinics have increased. The radiation effect from the use of low doses (0.005–0.200 Gy) in the field of diagnostics is still controversial. Recently, a study on the effects of the dose on survivors of atomic bomb radiation damage reported that the carcinogenic risk of a dose range of 0.0–0.1 Sv has been statistically verified [27]. It has been demonstrated that threshold values of radiation damage by the acute effects of low dose do not exist. The shielding method of low dose radiation is the best way of lead shielding. The polychromatic X-ray beam in CT scanning hardens because of the removal of low energy X-rays when they penetrate lead shielding. The beam hardening effect produces streak artifacts, which makes it difficult to observe surrounding tissues on the image [28]. However, bismuth shielding filters low energy X-rays and produces fewer beam hardening effects due to a metal artifact. As a result, the tissue that surrounds the shield can be observed. Bismuth shielding should be selectively located in order to avoid a significant degradation of the image quality. Image quality must be carefully evaluated when dose reduction methods are utilized. Kenneth D. Hopper *et al.* have reported that eye shielding in adult patients decreased the radiation dose by 39.6%–52.8% when bismuth thickness is changed. But when using 3.00 mm bismuth, it was difficult to observe the eye due to a beam hardening artifact, in spite of the decreased radiation dose [12]. Srinivasan Mukundan *et al.* have reported that the absorbing dose for the eye decreased about 39%, from 46 mGy to 28 mGy, when CT scanning of children's head phantom was done with the use of bismuth. They also stated that the image quality did not degrade [13].

The current study showed that the thicker the DRFS thickness, the more the radiation dose decreased, but the thicker DRFS thickness, the more that CT number values and noise values were increased by the beam hardening effect. However, the findings in this study suggested that CT number values were similar between 1.00 mm bismuth and 0.25 mm DRFS, and noise values were not significantly different among 1.00 mm bismuth, 0.15 mm DRFS, and 0.25 mm DRFS. This was because the radiation shielding was more effective as the nano-barium sulfate filled the gaps between the large oxide bismuth particles. Barium sulfate has been used as oral contrast media in patients in the department of radiology and has been harmless to the human body. Oxide bismuth is also a safe material, as it is chemically a very stable substance. The radiation shielding effect of bismuth is larger than that of barium, because bismuth's atomic number ($z=83$) is higher than barium's ($z=56$), which means that there is a much greater probability of the diagnostic X-ray photoelectric effect. There could be a more effective X-ray shielding for which the radiation reduction effect of bismuth could be the complement, because the added nano-barium particles correspond to the K-edge energy of bismuth in X-ray energy.

Bismuth shields are easy to use to protect critical organs located near the surface. These critical organs have a curved shape; therefore, shielding must be made thinner in order to contact the skin. However, conventional bismuth shields consist of bismuth mixed with latex, which is thicker than the DRFS. Our suggested DRFS shows a good adherence to curved surfaces because the material in the shields is thinner. We expect that DRFS can be used as a multi-type shield for curved surfaces.

In conclusion, the new DRFS shields can reduce doses to 15.61%, 23.05%, and 22.71% at 90 kVp, 120 kVp, and 140 kVp, respectively, more than reductions produced by conventional bismuth of the same thicknesses, while maintaining image quality. In addition, the DRFS can be made about 25% thinner than conventional bismuth. We conclude, therefore, that DRFS can replace the conventional bismuth and may be utilized as a new shield.

12

Table 1. Average dose reduction results with no shielding, bismuth shielding, and DRFS around the PMMA head phantom according to scanning condition of kVp.

| kVp | Shielding | Dose (mGy) | | Reduction from none (%) |
|---|---|---|---|---|
| | | Mean | Std. | |
| 90 | None | 9.25 | 1.19 | - |
| | Bismuth (1 mm) | 5.81 | 0.50 | 37.20 |
| | DRFS (0.15 mm) | 6.88 | 0.97 | 25.63 |
| | DRFS (0.25 mm) | 5.73 | 0.47 | 38.02 |
| | DRFS (1.00 mm) | 3.44 | 0.27 | 62.81 |
| | DRFS (2.00 mm) | 3.06 | 0.27 | 66.94 |
| 120 | None | 18.57 | 0.88 | - |
| | Bismuth (1 mm) | 12.99 | 0.76 | 30.02 |
| | DRFS (0.15 mm) | 16.05 | 1.58 | 13.56 |
| | DRFS (0.25 mm) | 12.84 | 0.75 | 30.85 |
| | DRFS (1.00 mm) | 8.71 | 0.63 | 53.07 |
| | DRFS (2.00 mm) | 6.88 | 0.38 | 62.95 |
| 140 | None | 28.28 | 2.36 | - |
| | Bismuth (1 mm) | 20.33 | 1.36 | 28.10 |
| | DRFS (0.15 mm) | 22.55 | 2.28 | 20.26 |
| | DRFS (0.25 mm) | 19.87 | 0.97 | 29.72 |
| | DRFS (1.00 mm) | 13.91 | 0.58 | 50.81 |
| | DRFS (2.00 mm) | 10.47 | 0.44 | 62.97 |

**Legends of Figures**

Figure 1. Setup of ionization chamber with PMMA head phantom and thyroid shielding for CT dose measurement.

Figure 2. Photograph of DRFS with 0.15 mm, 0.25 mm, 1.00 mm, and 2.00 mm thickness.

Figure 3. Cross-sectional schematic diagram of the DRFS: (a) coating fabric layer, (b) bismuth oxide power, and (c) nano-particles of barium sulfate in the liquid silicone rubber base.

Figure 4. Setup of PMMA head phantom with the center of the CT gantry, (a) with bismuth shielding, and (b) with DRFS around the PMMA head phantom.

Figure 5. The regions of interest (ROI), circles of about 300.0 ± 1.5 mm$^2$, were located at 3.0 ± 0.03 mm distance areas under the surface of the PMMA head phantom and at right and left areas at a distance of 3.0 ± 0.03 mm from ion chamber holes located at directions of 12 o'clock from the center of the PMMA head phantom.

Figure 6. CT transverse images of the PMMA head phantom at 120 kVp (ww/wl = 4095/1048), (a) without shielding and with (b) 1.00 mm bismuth shielding, (c) 0.15 mm DRFS shielding, (d) 0.25 mm DRFS shielding, (e) 1.00 mm DRFS shielding, and (f) 2.00 mm DRFS shielding.

Figure 7. (a) CT number and noise measurement (HU) at ROIs at under-surface right and left areas from ion chamber holes located at directions of 12 o'clock from the center of the PMMA head phantom.

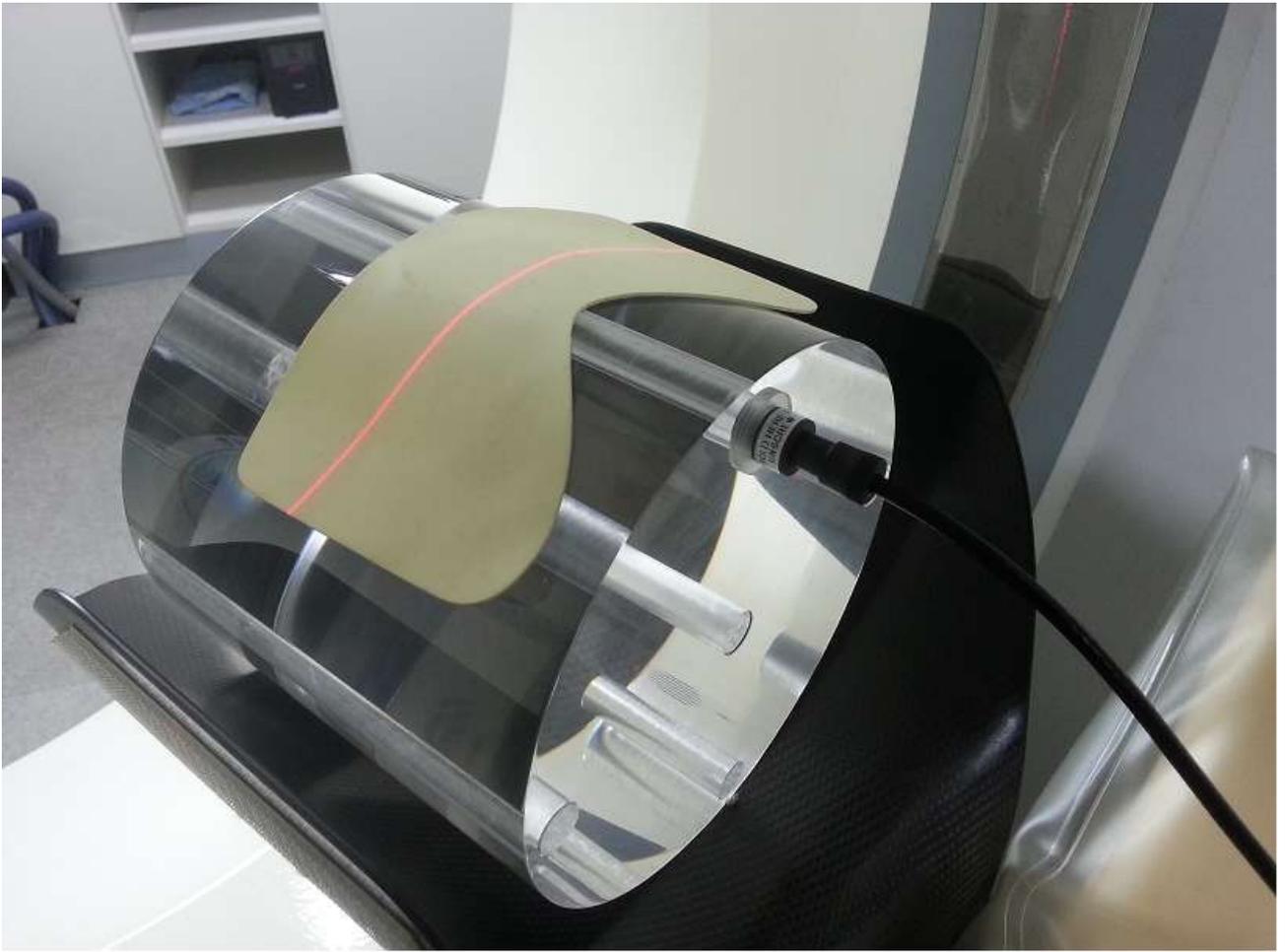

Figure 1

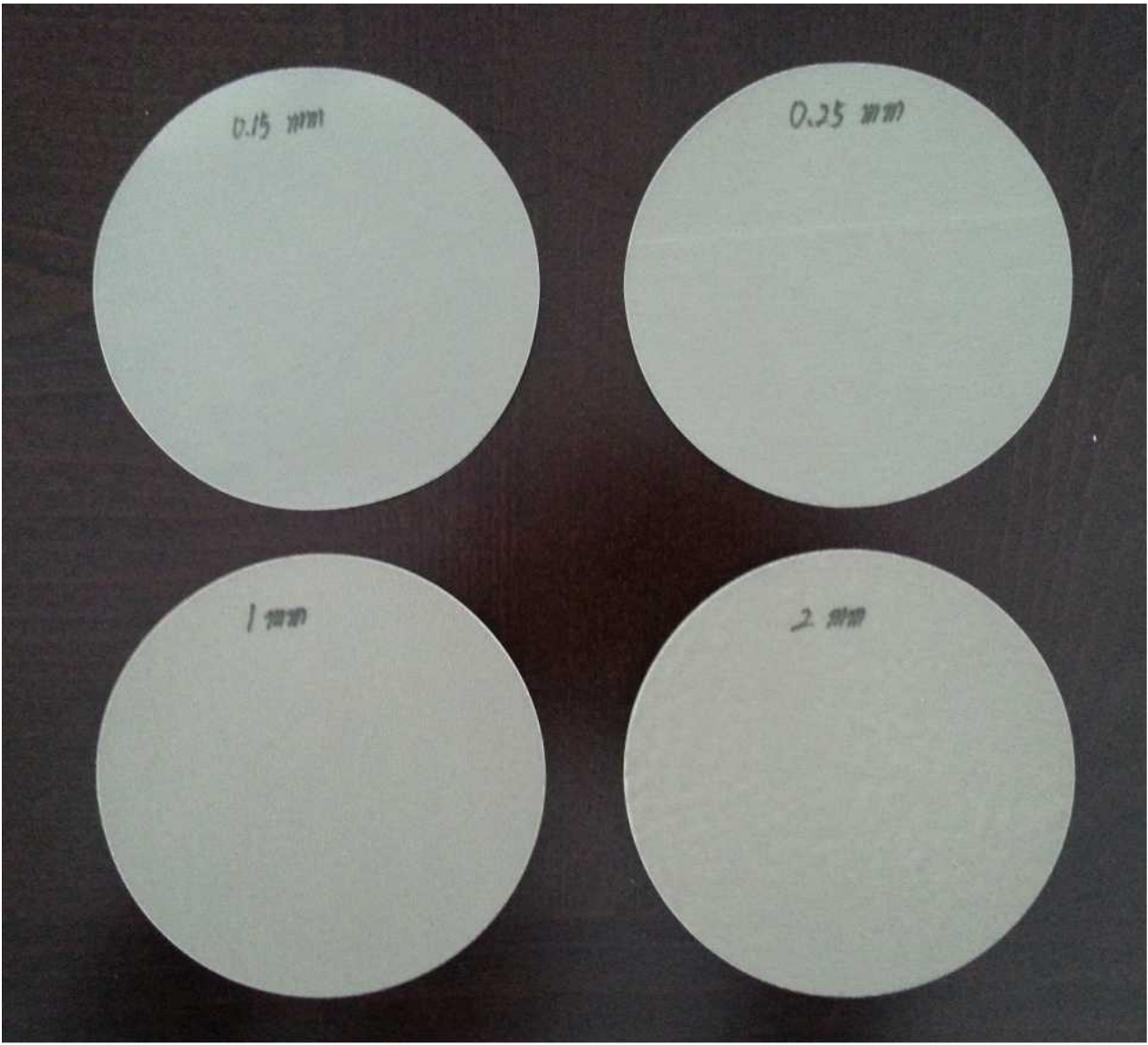

Figure 2

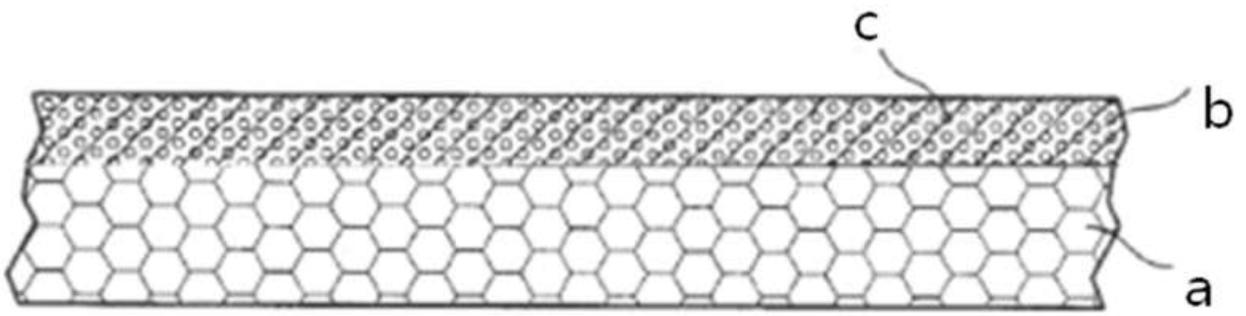

Figure 3

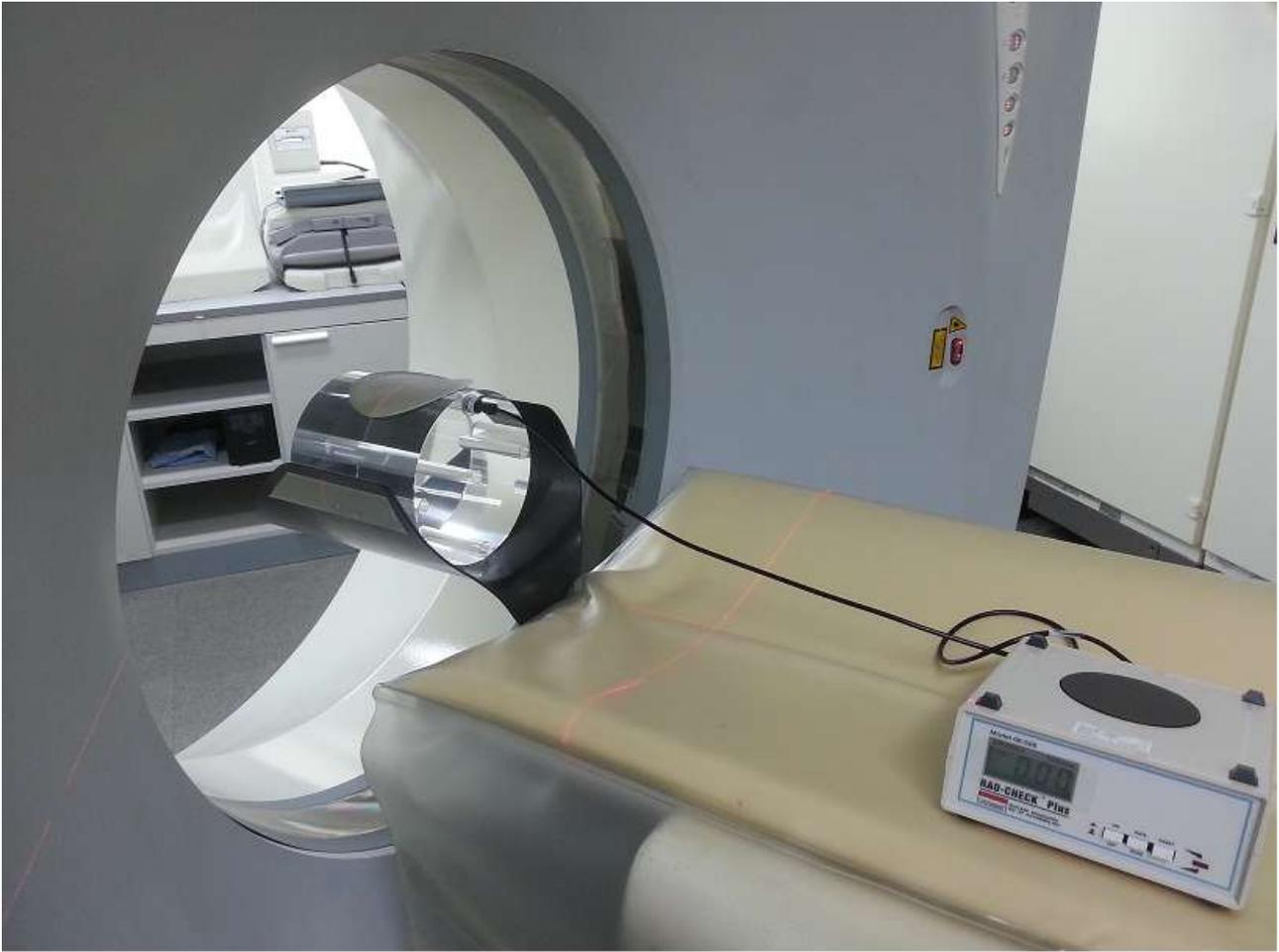

Figure 4 (a)

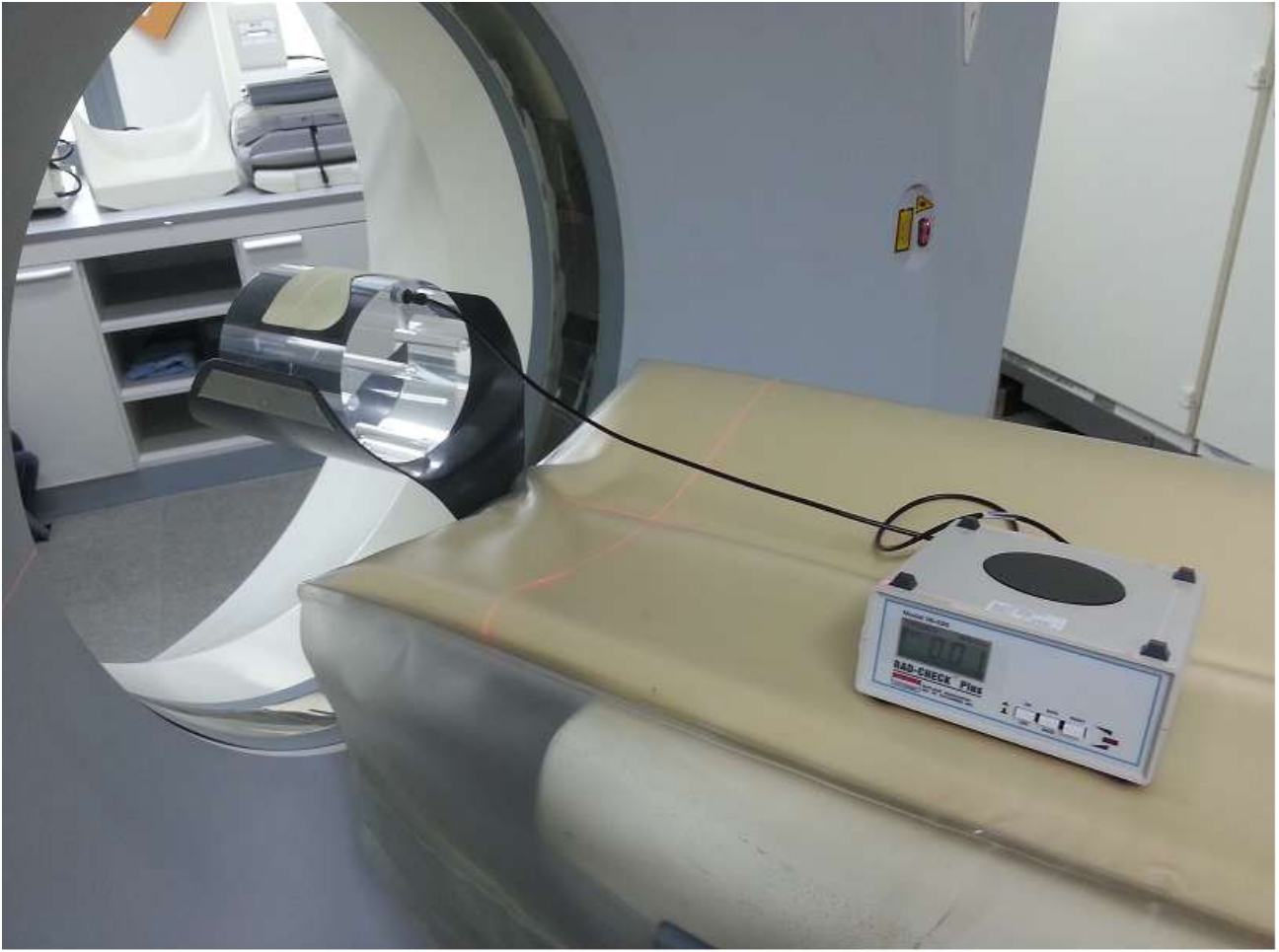

Figure 4 (b)

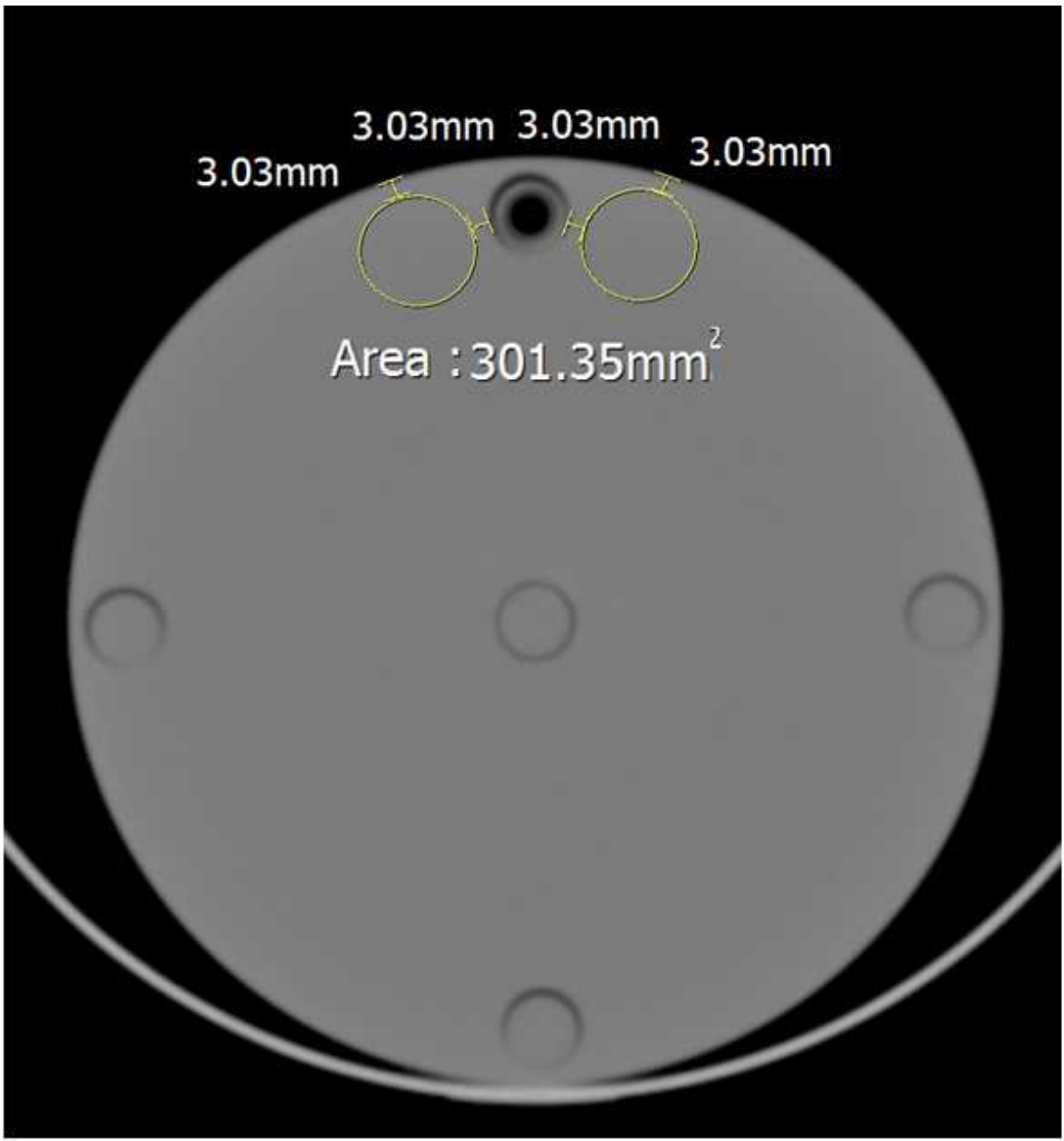

Figure 5

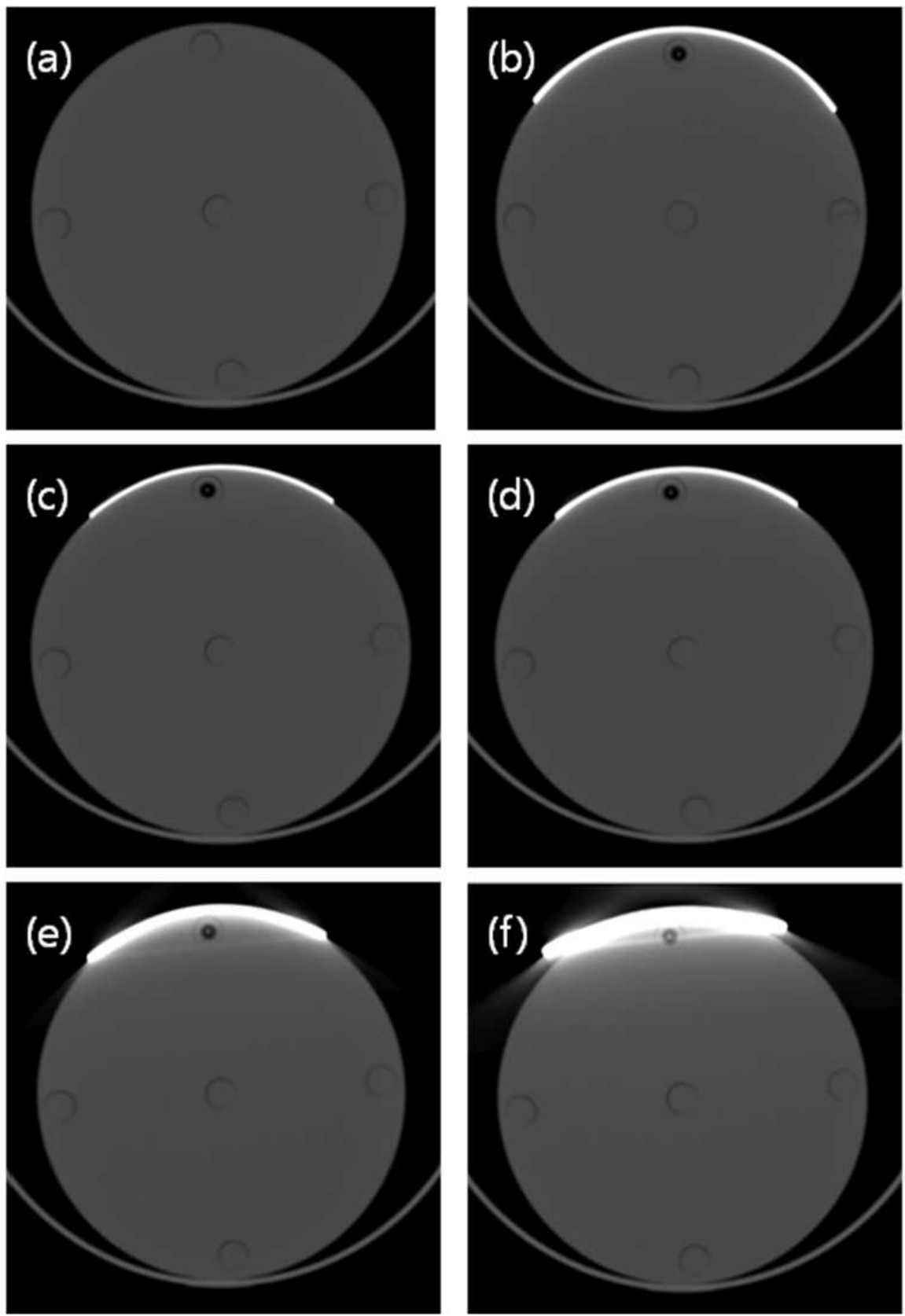

Figure 6

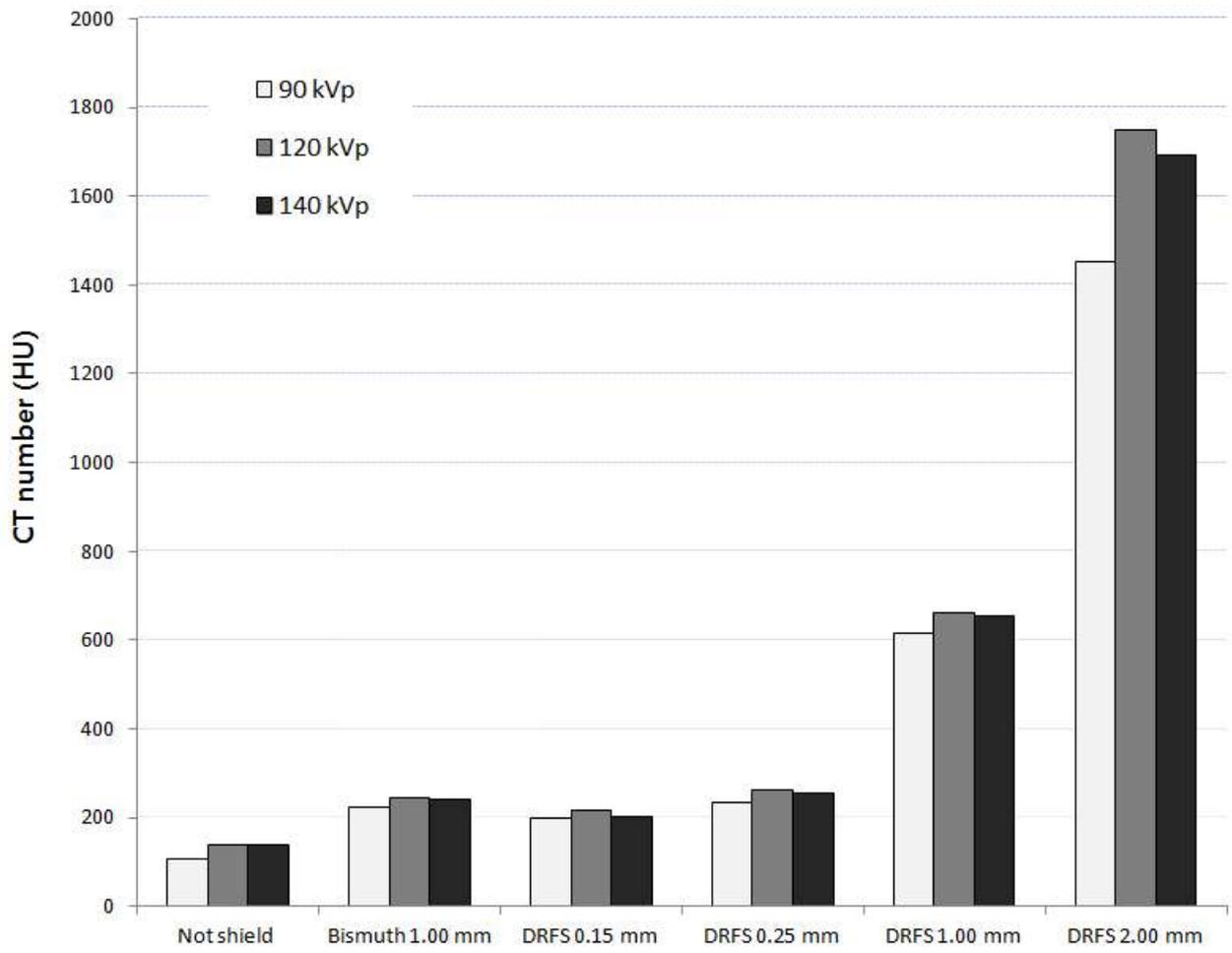

Figure 7 (a)

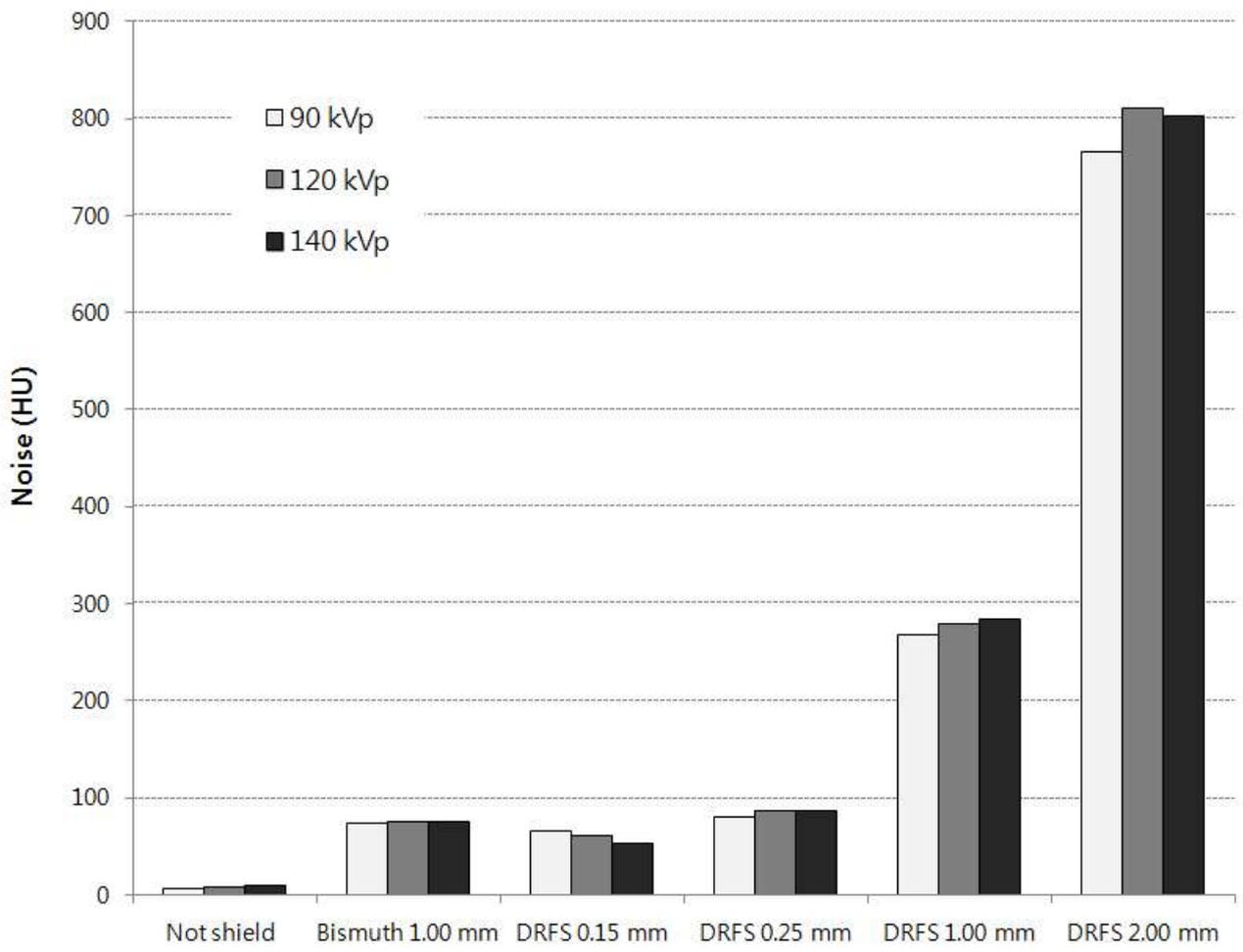

Figure 7 (b)